\documentclass[12pt]{iopart}

\usepackage{verbatim} 

\usepackage{graphicx}
\usepackage{float}
\restylefloat{figure}

\usepackage{amssymb,latexsym}
\usepackage{amsfonts}
\usepackage{amsthm}
\usepackage{stmaryrd}
\usepackage{color}

\begin{document}

\title{Entropy and efficiency of a molecular motor model}
\author{Eliran Boksenbojm and Bram Wynants}
\address{Instituut voor Theoretische Fysica, K.U.Leuven, 3001, Belgium.}
\eads{\mailto{Eliran.Boksembojm@fys.kuleuven.be}, \mailto{Bram.Wynants@fys.kuleuven.be}}

\begin{abstract}
In this paper we investigate the use of path-integral formalism and the concepts of entropy and traffic in the context of molecular motors.
We show that together with time-reversal symmetry breaking arguments one can find bounds on efficiencies of such motors.
To clarify this techinque we use it on one specific model to find both the thermodynamic and the Stokes efficiencies, 
although the arguments themselves are more general and can be used on a wide class of 
models. We also show that by considering the molecular motor as a ratchet, one can find additional bounds on the thermodynamic efficiency. 
\end{abstract}

\section{Introduction}

Understanding the functioning of living organisms has always been one of the most challenging
and most interesting aspects of science. It has been known for a while now that molecular motors 
play an important role in the working of these organisms \cite{sw,ll}. The term is actually used for a whole 
collection of small machines. However, they all have in common that they transform chemical energy into work,
and that thermal fluctuations are essential to their working. The energy needed comes mainly from the
hydrolysis reaction of ATP $ \leftrightarrow$ ADP + P, see e.g. \cite{mpb,mw,Ast}.\\ 

\noindent The chemical reaction cycle that drives the motor, is itself driven by maintaining a high concentration 
of ATP in the motor's environment, and is thus essentially out of equilibrium. Furthermore, the fact that molecular motors are very small, 
means that we are dealing with mesoscopic systems for which the dynamics involve stochastic elements.  Computations of e.g. efficiency 
of these molecular motors must therefore be quite different 
from their thermodynamic analogues.  The question of efficiency is indeed important and has been widely discussed 
already in the literature, see \cite{zha,kin,ebm,fld,emm,pjap}. 

\noindent In this paper we present a general method for obtaining bounds of efficiencies based on the 
relation between time-symmetry breaking and entropy production. These results are quite general and 
do not depend specifically on a model, even though we will illustrate the argument by using one specific model. 
We give bounds on both the usual thermodynamic and the Stokes efficiency.
Molecular motors are part of a wider class of systems called Brownian motors or ratchets \cite{drm,abm,srt}
We will use this relation between molecular motors and ratchet dynamics to derive other, less explicit, but possibly
better bounds for the efficiency. Also here we show that general probabilistic arguments can replace the 
thermodynamic procedure of computing efficiencies once there is a clear identification of the path-dependent 
entropy fluxes.\\

\noindent A well-known example of a molecular motor is kinesin, which moves cargo inside the cell away from the nucleus by ``walking'' 
along microtubules, with a step size of 8nm and loads of the order of pN, multiplying to work of the order 
of $k_B T$ at room temperature \cite{mpb}. This is a small scale for which a stochastic description is best fitting.\\
The hydrolysis reaction mentioned above will be modelled by a chemical cycle, described by a 
Markov jump process.
In a molecular motor this chemical cycle induces movement of a load in a spatial direction, thus doing work.
The spatial movement of the system will be described here by an overdamped Langevin equation, due to the viscosity
of the medium in which these motors live.
The coupling between the chemical cycle and the spatial movement is a so-called ``ratchet effect:'' to each 
chemical state of the system corresponds a different potential landscape, and by switching between these, a 
directed movement arises. 
The system is driven out of equilibrium. In the chemical state this is due to the maintained chemical potential, 
which is the actual energy input. The spatial component 
is in a sense also driven out of equilibrium by a non-gradient ``load''. However, the system will move 
in a direction opposite to this load, performing work as an output.

\noindent We start in section \ref{model} by describing the model and introducing some preliminary concepts. 
In section \ref{entropy_production} we will look at the entropy flux, the entropy production rate and the positivity 
of the entropy production using the path space formalism. In section \ref{efficiency} it will be shown that, 
in accordance with the second law of thermodynamics, both the thermodynamic efficiency and the Stokes efficiency 
are bounded. Section \ref{the_motor_as_a_ratchet} treats the motor as 
a ratchet. Using symmetry arguments from the ratchet formalism, we examine the efficiency more closely.\\

\noindent This paper is intended to illustrate two things: first of all a method to check the usual bounds on efficiencies - 
that they are smaller than 1. The model and the bounds on its efficiencies have been described already in \cite{zha}.
We focus here on the physical (model independent) argument, which revolves around the path-space formalism, entropy production and the 
second law of thermodynamics. 
Secondly, we illustrate how the ratchet formalism can produce possibly better bounds. In the model we use to illustrate the method,
the new bound is given by (\ref{newbound}).\\
 \\

\section{The model}
\label{model}

We consider a model for a molecular motor, coupled to a thermal bath at inverse temperature $\beta$, 
which has 2 degrees of freedom : a position $x$ on a one-dimensional circle with length $L$, modelling a periodic
potential landscape, and a chemical state $i\in\{1,\ldots,N\}$ with periodic boundary conditions: $N+1\equiv1$.
As time evolves, the system will therefore cover a trajectory (path) in the space of possible configurations.
Such a path starting at time zero and ending at a later 
time T will be denoted by $\omega = (x_t, i_t)_{0 \leq t \leq T}$.\\
 
The dynamics in the spatial direction x is taken to be an overdamped diffusion process 
(defined with the It\^o convention):
\begin{equation}
\label{Langevin}
 dx_t = \left[ \frac{f(x_t)}{\xi} - \frac{\phi_0'(x_t, i_t)}{\xi} \right] dt + \sqrt{2D} dB_t
\end{equation}
where $\xi$ is the viscous drag coefficient, $D$ is the free diffusion coefficient, related by the Einstein 
relation to the inverse temperature: $\beta = \frac{1}{\xi D}$. Furthermore $f$ is an external load, which is a
nonconservative forcing,  and $\phi_0(x_t, i_t)$ is a potential, which depends on the chemical state the motor is in. 
The prime denotes here the derivative with respect to x. Finally, $dB_t$ is standard Gaussian white noise.\\
 \\
The dynamics of the chemical state is governed by the transition rates $k_x (i, i+1)$ and $k_x (i, i-1)$, 
which depend on the position on the circle. Here we impose the physical condition of local detailed balance:
\begin{equation}
\label{rates}
 \frac{k_x (i,i+1)}{k_x (i+1,i)} = e^{ \beta \left[\phi_0(x,i) - \phi_0(x, i+1 )\right]+\beta\Delta\mu(i,i+1)}
\end{equation}
which just states that the rate for jumping to the right divided by the rate for jumping to the left is equal
to the exponential of the entropy produced by jumping to the right. Indeed, the entropy change in going from 
state $i$ to $i+1$ at position $x$ is $\beta$ times the heat flux between system and environment, which itself 
is equal to minus the change in energy ($\phi_0$) plus the chemical work $\Delta\mu(i,i+1)$. Note that
 $\Delta\mu(i,i+1) = -\Delta\mu(i+1,i)$. In general every full chemical cycle ($i\to i+N$) consumes a chemical 
energy $\Delta\mu = \sum_i\Delta\mu(i,i+1)$. When this $\Delta\mu$ is not zero, it is seen in 
(\ref{rates}) that detailed
balance is broken. The chemical potential thus provides the energy input for the motor, 
driving the chemical reaction cycle.\\

Let $p_t (x,i)$ be the probability density of finding the motor in position x in chemical state i at time t. 
From the dynamics described above, one can prove that this probability density evolves according to a set of 
coupled Fokker-Planck equations :
\begin{equation}
\label{fokpla} \frac{\partial p_t (x,i)}{\partial t} = -\frac{\partial J_{p_t}(x,i)}{\partial x}
+ I_{p_t} (x,i-1,i) - I_{p_t} (x, i ,i +1) 
\end{equation}
Where $J_{p_t}(x,i)$ is the probability current in the spatial direction for the i-th state 
and $I_{p_t} (x,i,i+1)$ is the probability current from state $i$ to state $i+1$.\\
These probability currents are given by
\begin{equation}
 I_{p_t} (x,i,i+1) = p_t (x,i) k_x(i,i+1) - p_t (x,i+1) k_x(i+1,i)
\end{equation}
\begin{equation}
 J_{p_t}(x,i) = \frac{1}{\xi} p_t(x,i)\left[ f(x) - \phi_0'(x,i) \right] - D p_t' (x,i) 
\end{equation}
Taking the left-hand-side of (\ref{fokpla}) zero gives the equation for the stationary measure, which we 
will henceforth denote by $\rho(x,i)$.
It is easily seen that when $f$ and the $\Delta\mu(i,i+1)$ are put to zero, the process satisfies detailed balance, 
with the equilibrium measure given by the Gibbsian distribution
\begin{equation}
 \rho_0 (x,i) = \frac{1}{\mathcal{Z}} e^{-\beta \phi_0 (x,i)}
\end{equation}
In the following it will be convenient to define such an equilibrium process as a reference process. 
More specifically with rates $k_{x_t}^{0}$ given by
\begin{eqnarray*}
 k_x^0(i,i+1) &=& k_x(i,i+1) e^{-\frac{\beta \Delta \mu(i,i+1)}{2}}\\
 k_x^0 (i+1,i) &=& k_x(i+1,i) e^{\frac{\beta \Delta \mu(i,i+1)}{2}}
\end{eqnarray*}
and with the same Langevin dynamics as in (\ref{Langevin}) with $f = 0$.

\section{Entropy Production}
\label{entropy_production}

\subsection{Path space formalism}
A useful way of describing a nonequilibrium system is via the distribution over its possible 
trajectories (paths), denoted by $P_{p_0}(\omega)$ with $p_0$ some initial measure. 
A consequence of the Markov property is that one can write that $P_{p_0}(\omega) = p_0(x_0,i_0)P(\omega|x_0,i_0)$,
where the second factor is now independent of the initial measure.
Such path-space distributions are best described as relative densities with respect to a reference process, 
denoted by $P^0$. For this purpose we take the reference equilibrium process as defined at the end of the last section. 
Given an initial distribution $p_0(x,i)$, the ratio of the probality of a path $\omega$ is given by
\begin{equation}
 \frac{dP_{p_0} (\omega)}{dP_{\rho_0}^{0} (\omega)} = \frac{p_0 (x_0, i_0)}{\rho_0 (x_0, i_0)}
\frac{P(\omega|x_0,i_0)}{P^0(\omega|x_0,i_0)}=:\frac{p_0 (x_0, i_0)}{\rho_0 (x_0, i_0)}e^{-A(\omega)}
\end{equation}
where $A$ is called the `action.' The action will be a sum of the action for a jump process (see e.g. Appendix 1 of \cite{kl}) and the 
action for a diffusion process (see e.g. \cite{mnw}): 
\begin{eqnarray}
 A(\omega)\label{actie} = & -\frac{\beta}{2} \int_{0}^{T} dx_t \circ f(x_t) + \frac{1}{2 \xi} \int_0^T dt \: f' (x_t)\nonumber \\ 
& - \frac{\beta}{4 \xi} \int_{0}^{T} dt \: f(x_t) \left[ f(x_t) - 2 \phi_0' (x_t, i_t) \right] \nonumber 
 - \sum\limits _{t \leq T} \log \left( \frac{k_{x_t} (i_{t^-}, i_t)}{k_{x_t}^{0} (i_{t^-}, i_t)} \right)\\
& + \int_{0}^{T} dt \sum\limits _j\left[ k_{x_t} (i_t,j) - k_{x_t}^{0} (i_t,j) \right]
\end{eqnarray}
where the first term of the action is a Stratonovitch-type stochastic integral, and $\sum_{t\leq T}$ means 
the sum over all the times that the system jumps to another chemical state. By $i_{t^-}$ and $i_{t}$ we denote 
respectively the chemical state before the jump and after the jump.\\

With the path space distribution, one can compute expectation values of observables. When started from an initial measure $p_0$, the expectation 
value of an observable $O(\omega)$ is denoted by
\[ \left<O(\omega)\right>_{p_0} := \int dP_{p_0}(\omega)O(\omega) \]
where the integral means the integral over all possible paths.

\subsection{Entropy flux}
It is well known (\cite{tre}) that the time-antisymmetric part of the action of a stochastic process gives 
the entropy flux $S(\omega)$ between the system and its surroundings.
Let $\theta$ be the time reversal operator, acting on paths as $\theta\omega = (x_{T-t},i_{T-t})_t$. 
The entropy flux is then given by
\begin{eqnarray}
\label{entav}
 S(\omega) & = & A(\theta\omega) - A(\omega) \nonumber\\
& = & \beta \int_{0}^{T} dx_t \circ f + \sum\limits_{t \leq T}\left[ \log \left( \frac{k_{x_t}(i_{t^-}, i_t)}{k_{x_t}^0 (i_{t^-}, i_t)} \right)-
\log \left( \frac{k_{x_t}(i_{t}, i_{t^-})}{k_{x_t}^0 (i_{t}, i_{t^-})} \right)\right]\nonumber\\
&=&  \beta \int_{0}^{T} dx_t \circ f +  \beta\sum\limits_{t \leq T}\Delta\mu(i_{t^-},i_t)
\end{eqnarray}
The two terms in this path-dependent entropy production already have a clear meaning. 
The first term comes from the motion in the spatial direction. In fact this first term 
is the entropy flux one would see, if the motor was only allowed to move in the spatial 
direction (for example by setting all rates $k$ to zero). Vice versa, the same is true 
for the second term, which comes from the motion in the chemical direction. We see that 
one full chemical cycle gives an entropy exchange between system and environment of $\beta\Delta\mu$, as expected.\\
The path space average of the entropy flux, starting with an initial distribution $p$, can be computed, again through the theory of Markov jump 
processes and diffusions:
\begin{eqnarray}
\label{entflux1}
 \left< S(\omega) \right>_{p_0} =& \beta  \int_{0}^{T} dt \int_0^L dx \sum\limits_{i} J_{p_t}(x, i)f(x)\nonumber \\
& + \beta\int_{0}^{T} dt \int_0^L dx \sum\limits_{i} I_{p_t} (x, i, i+1) \Delta \mu(i,i+1)
\end{eqnarray}

\subsection{Entropy production rate}
The total entropy production of the universe $S_u$ caused by this process is not only the entropy flux, 
but also the entropy change of the system itself. The entropy of the system is given by the relative entropy 
of the measure it is in, relative with respect to the equilibrium reference measure:
\begin{equation}\label{shannon}
 s(p) = -\sum_{i=1}^N\int_0^L dx \: p(x,i)\log \frac{p(x,i)}{\rho_0(x,i)}
\end{equation}
The change in the entropy of the system is then given by
\begin{equation}
\label{entchange}
 s(p_T) - s(p_0)
\end{equation}
Summing the two contributions ((\ref{entflux1}) and (\ref{entchange})) to the total entropy change $S_u$, 
it is then most elegantly written as
\begin{equation}
S^T_u(p_0) =  -\left<\log\frac{dP_{p_T} (\theta\omega)}{dP_{p_0}(\omega)}\right>_{p_0}
\end{equation}
where $p_T$ is the time-evolved measure determined by the Fokker-Planck equations (\ref{fokpla}).
The Markov property of our process makes that we can write the entropy production as 
$S^T_u(p_0) = \int_0^T dt \: \sigma(p_t)$ where $\sigma(p_t)=\lim\limits_{T \downarrow t}\frac{1}{T-t}S^{T-t}_u(p_t)$ 
is called the entropy production rate, straightforwardly computed to be 
\begin{eqnarray}
\label{epr}
 \sigma(p)   =&  \beta \xi \int dx \sum\limits_{i} \frac{J_{p}^{2} (x,i)}{p (x,i)}\\ & + 
\beta \int dx \sum\limits_{i} I_{p}(x,i,i+1) \left[ \Delta \mu (i, i+1) + \Phi(x,i)-\Phi(x,i+1) \right]\nonumber 
\end{eqnarray}
where $\Phi_{p}(x,i) = \phi_0(x,i) + \frac{1}{\beta} \log \, p (x,i)$.\\
Just as for the entropy flux, this entropy production rate consists of two separate terms, 
each having a clear meaning. For the first term, think of a new dynamics in which all chemical 
transition rates are set to zero. Then there is only motion in the spatial direction, and the 
entropy production for this process would exactly be the first term in (\ref{epr}). Suppose 
instead that we forbid all motion in the spatial direction, by letting $\xi\to\infty$. Then 
only the chemical reactions determine the process, and the entropy production for such a 
process is exactly the second term in (\ref{epr}). We will therefore write
\begin{equation}\label{epr2}
 \sigma(p) = \sigma_D(p) + \sigma_J(p)
\end{equation}
to denote that the entropy production is the sum of the entropy production of the diffusion process 
and of the jump process.

\subsection{Positivity of the entropy production}\label{posent}
The definitions of entopy flux and entropy production rate through path space measures 
have the advantage that one can easily prove their positivity: because of normalization 
of the path space measure we have for any measures $p,q$ and time $T$, and actually for 
any (well-behaved) dynamics:
\begin{equation}
\left< e ^{-\log\left( \frac{dP_{p}(\omega)}{dP_{q}(\theta\omega)}\right)}\right>_{p}=
\left<\frac{dP_{q}(\theta \omega)}{dP_{p}(\omega)}\right>_{p} = 1
\end{equation}
Jensen's inequality then dictates that
\begin{equation}
 \left<\log\left( \frac{dP_{p}(\omega)}{dP_{q}(\theta \omega)} \right) \right>_{p}\geq 0
\end{equation}
Using the definition of the entropy production rate $\sigma$ we find
\begin{equation}
\label{eprgeq1}
 \sigma (p) \geq 0
\end{equation}
Which is in essence the second law of thermodynamics: change of total entropy of the world is positive.
As said before, this entropy production rate can be written as a sum of an entropy production rate due 
to the diffusion in the spatial direction ($\sigma_D$) and an entropy production rate due to the jump 
process in the chemical direction ($\sigma_J$). As the argument of positivity above is quite general 
and does not depend on the specific dynamics of the system, it is also valid for dynamics where we set 
some rates to zero or let $\xi\to\infty$. In particular this means that the two terms of (\ref{epr2}) 
are separately positive
\begin{eqnarray*}
 \sigma_D (p) & \geq & 0 \\
\sigma_J (p) & \geq & 0 
\end{eqnarray*}

\section{Efficiency}
\label{efficiency}

To prove that the thermodynamic efficiency is bounded by one is not difficult. One just has to
realize that the total stationary entropy production of the process is the heat flux between the
system and the environment. This heat flux consists of the energy going into the system $Q_{in}$,
in our case through the chemical reaction cycle, and the power output $W_{out}$ by moving a load.
The fact that the stationary entropy production has to be positive then immediately gives that
$Q_{in}-W_{out}\geq 0$, or equivalently that the efficiency $\frac{W_{out}}{Q_{in}}\leq 1$.\\

But there is more. First of all, one has to realize that molecular motors like the one described here, are essentially
two processes which are coupled. The entropy production of the whole process is then the sum of the entropy
productions of the two separate processes. This is just a consequence of the Markov property. There is however one 
important difference: the stationary measure of the whole process is not the product of the stationary measures of 
the separate processes. This has as a consequence that the stationary entropy production of the whole process is rather 
the sum of transient entropy productions of the separate processes. Still, these entropy productions are both
positive, which leads to the bound on the Stokes efficiency, as illustrated for our model below.

\subsection{Thermodynamic efficiency}
Here we will work in the stationary case, i.e. the measure will be chosen to be $p_t = \rho$ 
the stationary measure. There is an immediate consequence of the (stationary) Fokker-Planck 
equation (\ref{fokpla}):
\begin{eqnarray*}
\frac{\partial V}{\partial x} &=& 0,\ \ \ \ \ \ \ \ \textrm{with}\ \ \ V := L\sum_iJ_{\rho}(x,i)\\
\nu(i) - \nu(i+1) &=& 0,\ \ \ \ \ \ \ \ \textrm{with}\ \ \  \nu := \int dx \: I_{\rho}(x,i,i+1) 
\end{eqnarray*}
which unambiguously defines the average velocity $V$  and the mean chemical reaction rate $\nu$. 
As usual, the efficiency is the (average) mechanical work that the motor can perform on the load, divided by 
the (average) chemical energy that is provided.
With the above definitions, the thermodynamic efficiency is then clearly equal to
\begin{eqnarray}
 \eta_T = -\frac{\bar{f} V}{\nu \Delta \mu}
\end{eqnarray}
where $\bar{f} = \frac{1}{L}\int_0^L dx f(x)$ is the average force on the circle. The minus sign in the definition is
there because for the motor to work, one needs that $\nu \Delta \mu>0$ and $\bar{f} V<0$.
 As we know, the second law of thermodynamics should give a bound 
on this efficiency, namely that it is smaller than $1$.
Indeed, in the stationary case the entropy flux (\ref{entflux1}) becomes
\begin{equation}\label{statentflux}
 \left< S(\omega) \right>_{\rho} = \beta T[ V\bar{f} +  \nu\Delta \mu]
\end{equation}
 When the system is in the stationary state, its Shannon entropy (\ref{shannon}) is constant in time, so that
\begin{equation}
 \left< S(\omega) \right>_{\rho} = T\sigma(\rho)\geq 0
\end{equation}
Together with (\ref{statentflux}), this gives that $\eta_T \leq 1$.

\subsection{Stokes efficiency}

In many cases the goal of a molecular motor is to carry cargo across a viscous environment 
(e.g. protein transport within a cell). In this case the useful power of the motor is just 
the viscous drag force times the average velocity: $\xi V\cdot V$. While the total power 
input from the environment is the chemical input through the chemical potential, and mechanical 
input through the external force $f$. Note that $f$ is now not regarded as a load, but as an external 
force. 
Because of this, the Stokes efficiency was introduced in (\cite{ste}) to have a more useful 
definition of efficiency than the previous thermodynamic one:
\begin{equation}
 \eta_s = \frac{\xi V^2}{\bar{f}V + \nu\Delta\mu}
\end{equation}
We will show here that for this model, the Stokes efficiency is bounded by one, 
just as the thermodynamic efficiency. Again, this will follow from the second law 
of thermodynamics, by considering the entropy production rate $\sigma(\mu)$ as the 
sum of the two entropy production rates $\sigma_D$ and $\sigma_J$ for the mechanical 
and the chemical cycle. We already proved that these two quantities are positive 
(see section \ref{posent}). Note however, that these two quantities are entropy productions 
for restricted dynamics, one where the chemical movement is prohibited and the other where 
mechanical movement is not possible. This also means that the stationary measure of the 
molecular motor is not the stationary measure for these restricted processes. Luckily, 
this poses no problem, because the positivity of the entropy was proven for any measure $p$. 
We thus have that $\sigma_D(\rho)\geq 0$ and $\sigma_J(\rho)\geq 0$. We find then
\begin{equation}
 1\geq\frac{\sigma_D (\rho)}{\sigma_D(\rho) + \sigma_J(\rho)} =\frac{\beta \xi 
\int dx \sum\limits_i \rho(x,i)\left(\frac{J_{\rho}(x,i)}{\rho(x,i)}\right)^2}
{\beta \left[ f_{ext}V + \nu \Delta \mu \right]}
\end{equation}
This inequality leads to the bound on the Stokes efficiency:
\begin{eqnarray*}
 \eta_S & = & \frac{ \xi V^2}{ f_{ext} V + \nu \Delta \mu} = \frac{\xi 
\left[ \sum\limits_{i=1}^{N} \int J_{\rho}(x) dx \right]^2 }{f_{ext} V + \nu \Delta \mu}  \\
& \leq  & \frac{\beta \xi \int dx \sum\limits_i \frac{J^2_{\rho}}{\rho}}{\beta 
\left[ f_{ext}V + \nu \Delta \mu \right]} \\
& \leq & 1
\end{eqnarray*}
where the first inequality is an immediate consequence of the positivity of the variance of $\frac{J_{\rho}}{\rho}$.

\section{The motor as a ratchet}
\label{the_motor_as_a_ratchet}

In this section we will show that the molecular motor as modelled here is a exactly a ratchet as 
defined in \cite{drm}. We will then use properties of ratchets to derive another,
possibly better, bound on the thermodynamic efficiency\\

\subsection{Another symmetry}
The quantity that determines the power output of the motor, is the current in the spatial direction.
For a specific trajectory $\omega$, this current $j_T$ is defined by the following relation:
\begin{equation}
 \int dx g(x)j_T(x) := \int_0^Tdx_t\circ g(x_t)
\end{equation}
where again a stochastic integral is taken with the Stratonovich interpretation. With this definition, 
one can compute that $\left< j_T(x) \right>_{\mu}=\int_0^TdtJ_{\mu_t}(x)$.

It is obvious that $j_T$ is independent of the specific sequence of chemical states it went through, 
it only depends on the movements in the spatial direction. 
It will therefore be convenient to define a new symmetry operation $\Gamma$ as follows: 
$\Gamma\omega := (x_t,N+1-i_t)_t$. The current $j_T(x)$ then satisfies
\begin{eqnarray*}
 \Gamma j_T(x) &=& j_T(x)\\
\theta j_T(x) &=& -j_T(x)
\end{eqnarray*}
which makes it a ratchet current, exactly as defined in \cite{drm}.\\

We now have two symmetry operations: the time-reversal $\theta$, and the ``chemical reversal'' $\Gamma$.
With this, we can repeat the argument of the positivity of the entropy production, 
but now with our new symmetry operator. 
Let us therefore split the action (\ref{actie}) into four parts, according to these symmetries. 
The antisymmetric part of the action under time reversal was already seen to be the entropy production. 
The time-symmetric part of the action was also earlier examined in e.g. \cite{mncan,mnw}, and named traffic $\mathcal{T}$, so that
\[ A = \frac{1}{2} \left[\mathcal{T} - S\right] \]
Using the symmetry operator $\Gamma$, the action splits further as follows:
\[ A = \frac{1}{4}[\mathcal{T}_+ - \mathcal{T}_- -S_+ + S_-] \]
where the plus and minus denote respectively the symmetric and antisymmetric parts of the functions 
under $\Gamma$, e.g. $S_- = [\Gamma S - S]$. For the argument to work,
it is necessary that the reference equilibrium process is also symmetric under $\Gamma$. Even though it is always 
possible to find such a reference, to make computations simpler, we will assume that 
the reference process we defined earlier is already symmetric under $\Gamma$:
\begin{eqnarray*}
 \phi_0(x,i) &=& \phi_0(x,N+1-i)\\
 k_x^0(i,j) &=& k_x^0(N+1-i,N+1-j)
\end{eqnarray*}
\noindent With this, we repeat the argument of section \ref{posent}, but now for $\Gamma\theta$ instead of $\theta$. 
This argument gives here that
\[ \left< \log \frac{\rho(x_0, i_0)}{\rho(x_T, N+1-i_T)} \right>_{\rho} + 
\frac{1}{2}\left< S_+(\omega) + \mathcal{T}_-(\omega) \right>_{\rho} = 
\left<\log \frac{dP_{\rho}(\omega)}{dP_{\rho}(\Gamma\theta\omega)}\right>_{\rho}\geq 0 \] 
which means in particular that
\begin{equation}
\left< \mathcal{T}_{-}(\omega) \right>_{\rho} \geq -2 \left< \ln \frac{\rho (x_0, i_0)}{\rho(x_T, N+1-i_T)} 
\right>_{\rho} - 2 \beta T V \bar{f}
\end{equation}
As this equality is true for any time $T$, and as the first term on the right-hand side of this inequality is not 
extensive in time, one sees that
\begin{equation}
\label{bounde}
\frac{1}{T}\left< \mathcal{T}_{-}(\omega) \right>_{\rho} \geq  - 2 \beta V \bar{f}
\end{equation}
We see therefore that the time-symmetric part of the action plays an important 
role in ratchets and thus in molecular motors. Let us therefore try to compute $\mathcal{T}_-$ for our model.

\subsection{The efficiency revisited}
\label{stronger_bound_on_the_efficiency}

In this case it is easily seen that 
\[ T_-(\omega) = 2 \int_{0}^{T} dt \sum_j\left[ k_{x_t} (N+1-i_t,j) - k_{x_t} (i_t,j) \right] \]
After a straightforward calculation, one finds
\begin{eqnarray}
 \left< \mathcal{T}_{-}(\omega) \right>_{\rho} =& 2 T \sum\limits_i \left[ \cosh\left(\frac{\beta}{2} \Delta \mu_{-}(i,i+1)\right) - 1 \right] 
\tau(i,i+1)\nonumber \\ & + 2 \nu T \sum\limits_{i} \sinh \left(\frac{\beta}{2} \Delta \mu_{-}(i,i+1)\right)
\end{eqnarray}
where we have defined $\Delta\mu_{-}(i,i+1) = \Delta\mu(N+1-i,N-i)-\Delta\mu(i,i+1)$, and
\[ \tau(i,i+1) = \int_{0}^{L} dx \: \rho(x,i) k_x(i,i+1) +  \rho(x,i+1) k_x(i+1,i)  \]
is the symmetric counterpart of the current in the chemical direction.\\
It follows now from (\ref{bounde}) that
\begin{eqnarray}
 -\beta V \bar{f}\leq &\sum\limits_i \left[ \cosh\left(\frac{\beta}{2} \Delta \mu_{-}(i,i+1)\right) - 1 \right] \tau(i,i+1)\nonumber \\ & + \nu 
\sum\limits_{i} \sinh \left(\frac{\beta}{2} \Delta \mu_{-}(i,i+1)\right) 
\end{eqnarray}
We find a new boundary for the thermodynamic efficiency
\begin{eqnarray}\label{newbound}
 \eta_T \leq & \sum\limits_i \frac{\tau(i,i+1)}{\beta \nu \Delta \mu} \left[ \cosh\left( \frac{\beta}{2} \Delta \mu_{-}(i,i+1) \right) - 1 \right]
\nonumber \\ & + \frac{1}{\beta \Delta \mu} \sum\limits_{i} \sinh \left( \frac{\beta}{2} \Delta \mu_{-} (i,i+1) \right)
\end{eqnarray}
Even though this is not a very explicit formula, one can see the importance of fluctuations: the quantity
$\tau(i,i+1)$ deals with the average number of jumps between states $i$ and $i+1$, thus being a measure of the
total activity in the chemical direction. When $\tau(i,i+1)/\nu$ is big, then the efficiency of the motor may also be big.
Furthermore, this traffic plays a fundamental role in fluctuation theory, see e.g. \cite{mnw,mncan}.

\section{Conclusions}

We showed in this paper that the boundedness of the efficiencies (both thermodynamic and Stokes) 
for a large class of models for molecular motors can be deduced with model-independent arguments. 
These general probabilistic arguments are based on the path-space formalism and the relation between 
time-symmetry breaking and entropy production. By also showing that such a motor can be seen as a 
ratchet other boundaries were found for the thermodynamic efficiency.\\

To find more explicit results approximations should be made, or otherwise a numerical analysis.
This is beyond the scope of this text, e.g. in \cite{ebm,emm,pjap} such approximations and numerical analyses are made,
however by using a different method. Therefore it could be interesting for further research to make these approximations and/or
numerical analyses, starting from our method, to compare with the results in these papers.

\ack
Discussions with C. Maes and W. De Roeck are gratefully acknowledged.\\
B.W. is an aspirant of the Fund for Scientific Research - Flanders, FWO.
E. B. has benefited from K.U.Leuven Grant no. OT/07/034A.


\Bibliography{10}

\bibitem{sw}
M.~Schliwa and G.~Woehlke :
\newblock Molecular motors,
\newblock {\em Nature} \textbf{422}, 759--765 (2003).

\bibitem{ll}
R.~Lipowsky and S.~Liepelt :
\newblock Molecular motors and stochastic networks,
\newblock {\em Banach Center Publ.} \textbf{80}, 167--195 (2008).

\bibitem{mpb}
C.~Bustamante, Y.R.~Chemla, N.R.~Forde and D.~Izhaky :
\newblock Mechanical processes in biochemistry,
\newblock {\em Annu. Rev. Biochem.} \textbf{73}, 705--748 (2004).

\bibitem{mw}
C.~Maes and M.~van Wieren :
\newblock A markov model for kinesin,
\newblock {\em J. Stat. Phys.} \textbf{112}, 329--355 (2003).

\bibitem{Ast}
R.~D.~Astumian :
\newblock Thermodynamics and kinetics of a brownian motor,
\newblock {\em Science} \textbf{276}, 917--922 (1997).

\bibitem{zha}
Y.~Zhang :
\newblock The efficiency of molecular motors,
\newblock {\em J. Stat. Phys.} \textbf{134}, 669--679 (2009).

\bibitem{kin}
E.~Calzetta :
\newblock Kinesin and the crooks fluctuation theorem,
\newblock {\em Eur. Phys. J. B} \textbf{68}, 601--605 (2009).

\bibitem{ebm}
J.M.R.~Parrondo, J.M.~Blanco, F.J.~Cao and R.~Brito :
\newblock Efficiency of brownian motors,
\newblock {\em Europhys. Lett.} \textbf{43}, 248--254 (1998).

\bibitem{fld}
D.~Lacoste, A.W.C.~Lau and K.~Mallick :
\newblock Fluctuation theorem and large deviation function for a solvable model of a molecular motor, 
\newblock {\em Phys. Rev. E} \textbf{78}, 011915 (2008).

\bibitem{emm}
T.~Schmiedl and U.~Seifert :
\newblock Efficiency of molecular motors at maximum power,
\newblock {\em Europhys. Lett.} \textbf{83}, 30005 (2008).

\bibitem{pjap}
A.~Parmeggiani, F.~J\"{u}licher, A.~Ajdari and J.~Prost :
\newblock Energy transduction of isothermal ratchets: Generic aspects and specific examples close to and far from equilibrium,
\newblock {\em Phys. Rev. E} \textbf{60}, 2127 (1999)

\bibitem{drm}
W.~De Roeck and C.~Maes :
\newblock Symmetries of the ratchet current,
\newblock {\em Phys. Rev. E} \textbf{76}, 051117 (2007).

\bibitem{abm}
A.~Fiasconaro, W.~Ebeling and E.~Gudowska-Nowak :
\newblock Active brownian motion models and applications to ratchets,
\newblock {\em Eur. Phys. J. B} \textbf{65}, 403--414 (2008).

\bibitem{srt}
R.D.~Astumian :
\newblock Symmetry relations for trajectories of a brownian motor,
\newblock {\em Phys. Rev. E} \textbf{76}, 020102 (2007).

\bibitem{kl}
C.~Kipnis and C.~Landim.
\newblock {\em Scaling limits of interacting particle systems}.
\newblock Springer, 1999.

\bibitem{mnw}
K.~Neto\u{c}n\'{y}, C.~Maes and B.~Wynants :
\newblock Steady state statistics of driven diffusions,
\newblock {\em Physica A} \textbf{387}, 2675--2689 (2008).

\bibitem{tre}
C.~Maes and K.~Neto\u{c}n\'{y} :
\newblock Time-reversal and entropy,
\newblock {\em J.Stat. Phys.} \textbf{110}, 269--310 (2008).

\bibitem{ste}
H.~Wang and G.~Oster :
\newblock The stokes efficiency for molecular motors and its applications,
\newblock {\em Europhys. Lett.} \textbf{57}, 134--140 (2002).

\bibitem{mncan}
C.~Maes and K.~Neto\u{c}n\'{y} :
\newblock Canonical structure of dynamical fluctuations in mesoscopic nonequilibrium steady states,
\newblock {\em Europhys. Lett.} \textbf{82}, 30003 (2008).

\endbib
\end{document}